\documentclass[draft]{aipproc}
\usepackage{amsmath,amssymb,wasysym}
\usepackage{graphicx}

\layoutstyle{8x11single}

\begin{document}

\title{Moreton Waves and EIT Waves Related to the Flare Events of June 3, 2012 and July 6, 2012}


\author{Agustinus Gunawan Admiranto}{
  address={Space Science Center, National Institute of Aeronautics and Space (LAPAN)}
}

\author{Rhorom Priyatikanto}{
  address={Space Science Center, National Institute of Aeronautics and Space (LAPAN)},
  altaddress={Astronomy Program, Faculty of Mathematics and Natural Sciences, Institut Teknologi Bandung}
}

\author{Ubaydillah Yus'an}{
  address={Astronomy Program, Faculty of Mathematics and Natural Sciences, Institut Teknologi Bandung}
}

\author{Evaria Puspitaningrum}{
  address={Astronomy Program, Faculty of Mathematics and Natural Sciences, Institut Teknologi Bandung}
}

\begin{abstract}
We present geometrical and kinematical analysis of Moreton waves and EIT waves observed on June 3, 2012 and Moreton waves observed on July 6, 2012. The Moreton waves were recorded in H$\alpha$ images of Global Oscillation Network Group (GONG) archive and EIT waves obtained from SDO/AIA observations, especially in 193 nm channel. The observed wave of June 3 has angular span of about $70^{\circ}$ with a broad wave front associated to NOAA active region 11496. It was found that the speed of the wave that started propagating at 17.53 UT is between 950 to 1500 km/s. Related to this wave occurence, there was solar type II and III radio bursts. The speed of the EIT in this respect about 247 km/sec. On the other hand, the wave of July 6 may be associated to X1.1 class flare that occurred at 23.01 UT around the 11514 active region. From the kinematical analysis, the wave propagated with the initial velocity of about 1180 km/s which is in agreement with coronal shock velocity derived from type II radio burst observation, $v\sim1200$ km/s.
\end{abstract}

\maketitle

\section{Introduction}
Moreton and EIT waves are some kind of large scale disturbances which occur in the  atmosphere of the Sun. These waves are enigmatic phenomena which are somewhat related with solar energetic phenomena such as flare and coronal mass ejection, although the exact relations haven't discovered yet. Moreton wave can be observed in H$\alpha$ wavelength, so it is inferred that this wave occur in the chomosphere. This wave was discovered by Moreton \citep{moreton1960} and propagate with the velocity in the range of 1000-1200 km/second. 

On the other hand, the discovery of EIT wave was only made recently by the advent of space based observatory. The first observations of EIT waves was conducted using Extreme Ultraviolet Imaging Telescope \citep[EIT, see][]{dela1995} on-board the Solar and Heliospheric Observatory (SOHO). EIT waves propagate along the quiet Sun, sometimes covering most of the solar disk, but they never occur in the coronal holes or active regions. These waves also exhibit different characteristics with Moreton wave in the angular extent and its velocity. While Moreton wave propagate with speed of 1000-1500 km/sec \citep{athay1960}, the speed of EIT waves are much lower, in the order 100-250 km/sec, and sometimes lower than this value \citep{klassen2000}. The EIT waves propagate more isotropically than that of Moreton waves which have average angular extent of 90 degrees.

Nowadays, there are some disputes related to the nature of both waves, are they related at all? Are the physical mechanisms in both phenomena the same, only different in manifestations due to the different place of occurrence? Some authors even said that these waves are not waves at all, only some kind of shock which are caused by the flare or coronal mass ejection \citep{chen2011}. On the other hand, Grechnev et al. \citep{grechnev2011} investigated some flares and coronal mass ejections and Moreton and EIT waves and they concluded that there are some relations among Moreton and EIT waves with explosives phenomena in the solar atmosphere.

This research focuses on the Moreton and EIT waves which are related with flare events on June 3 and July 6, 2012. We try to investigate the relationship between the Moreton wave, EIT wave, and other phenomena like solar radio burst which was caused by shocks that trigger Moreton or EIT waves. 

\begin{figure}
\centering
\framebox{\includegraphics[width=0.9\textwidth]{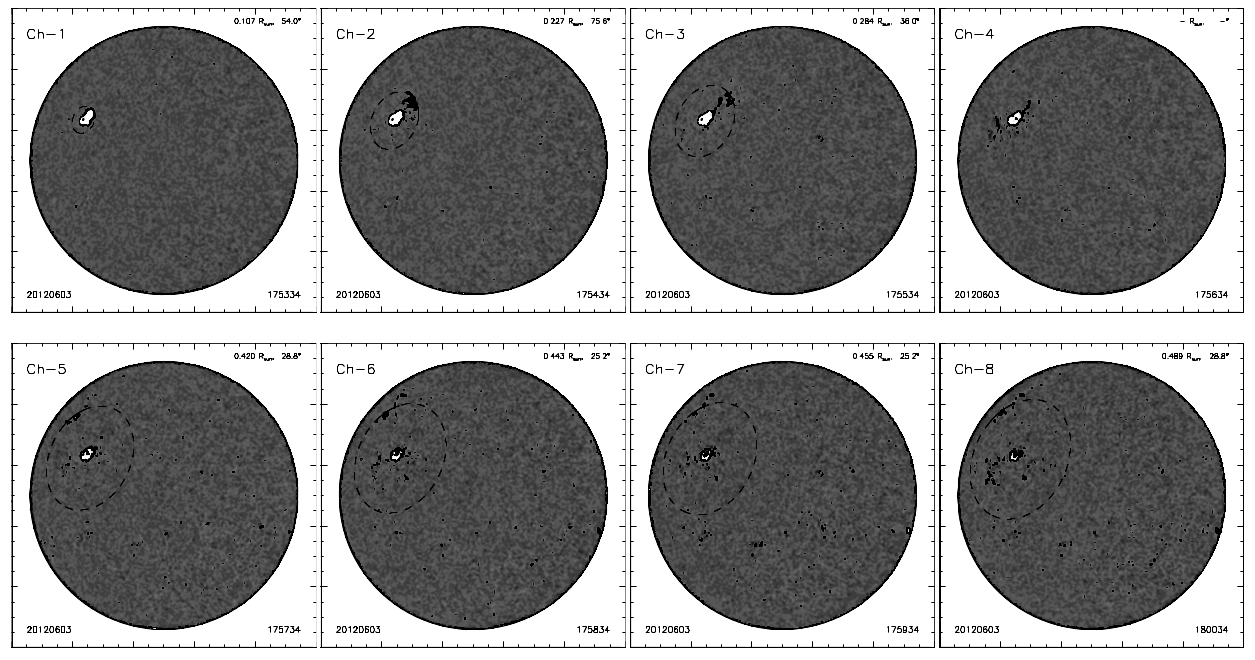}}
\caption{H$\alpha$ differential images related to the flare of June 3, 2012. These images reveal the propagation of shock waves to north-east and north-west direction.}
\label{halpha}
\end{figure}

\begin{figure}
\centering
\framebox{\includegraphics[width=0.9\textwidth]{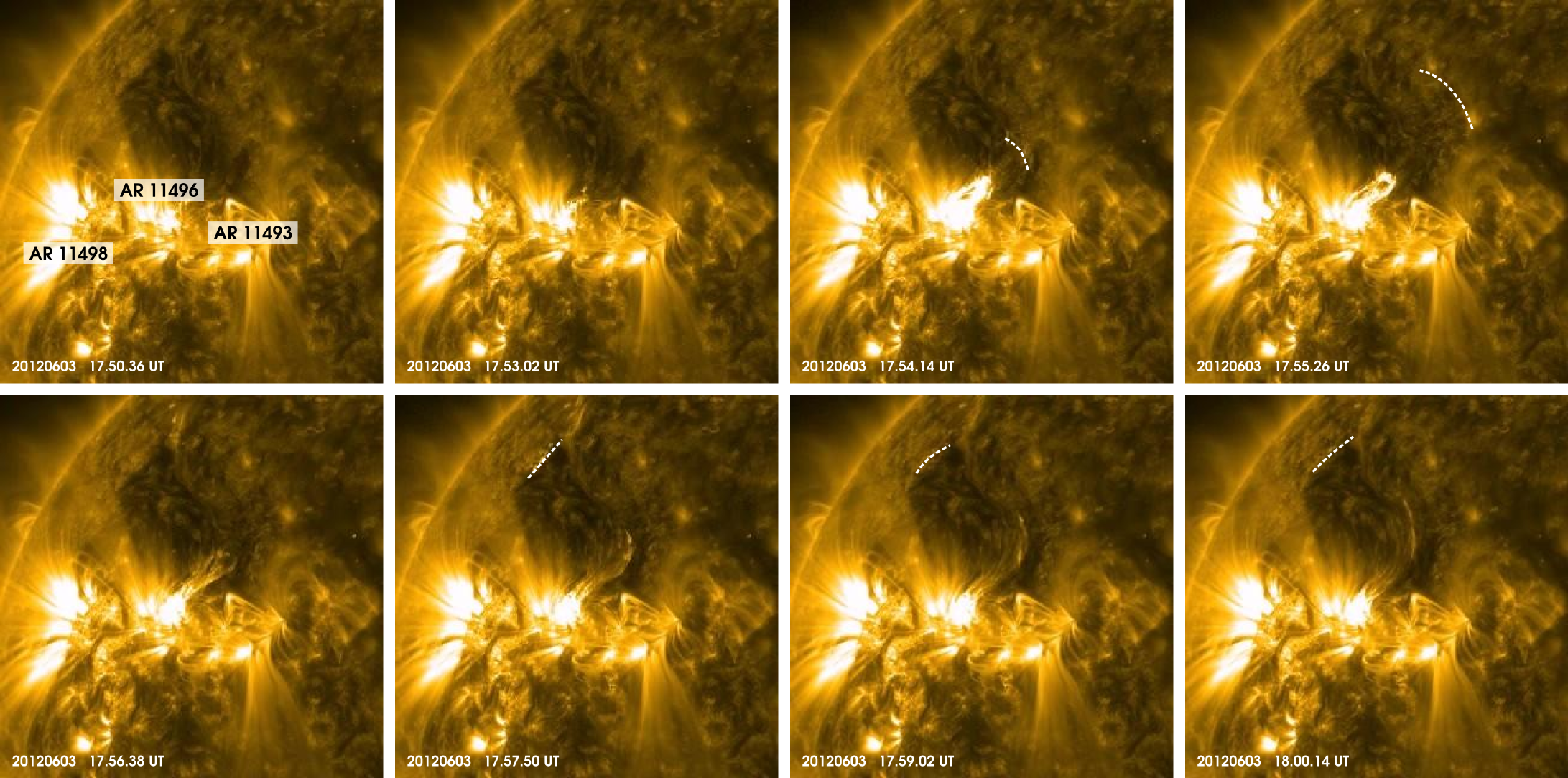}}
\caption{Zoomed images of the Sun at EUV ($\lambda=193$ nm) show similar shock wave (marked by dashed lines).}
\label{eit}
\end{figure}

\begin{figure}
\centering
\framebox{\includegraphics[width=0.7\textwidth]{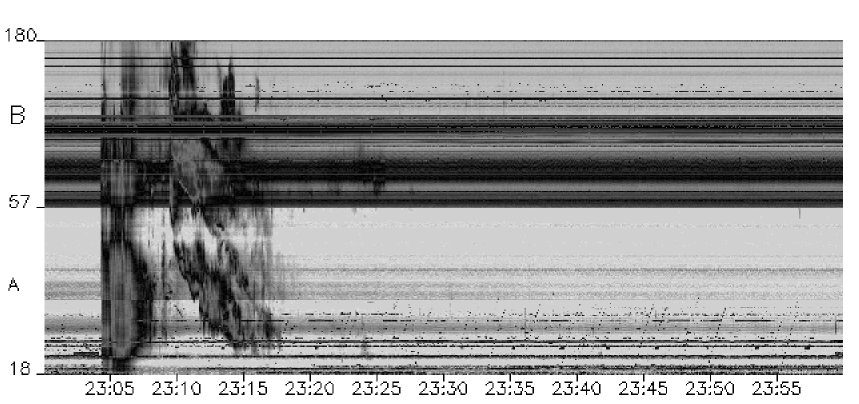}}
\caption{Radio spectrum (18 to 180 MHz) from Culgoora Observatory, Australia obtained in July 6, 2012. Type III burst occurred at 23.05, while type II burst started later at 23.10.}
\label{radio}
\end{figure}

\section{Data and Method}
This analysis used GONG (Global Oscillation Network Group) to get H$\alpha$ images data to infer the existence Moreton wave and SDO/AIA (Solar Dynamic Observatory/Atmospheric Imaging Assembly) data to get the EUV data. Solar radio burst data was obtained from \url{ips.gov.au}.

The first event occured in June 3, 2012 between 17:52 -- 17:57 UT after M3.3 flare in the active region NOAA 11496. During that time, four among six GONG observatories (Big Bear, Mauna Loa, Cerro Tololo, Teide) were able to record the central line H$\alpha$ images with time resolution of about 20 seconds. For the second event of July 6, there was an X1.1 flare in the active region NOAA 11515 occurred at 23.01 UT. H$\alpha$ images related to the event of July 6 are obtained from Mauna Loa observatory.

EUV images at $\lambda=193$ nm from SDO/AIA revealed coronal shock waves that related to the same event of June 3, 2012 (Figure \ref{eit}). One can see the movements of the shock waves and from these movements one can determine the speed of the wave's propagation. For the second event, we could not detect any clear indication of coronal wave since the event occured near the limb. Besides, the lower time resolution of the acquired images hinder us to find the wave.

Visual inspection was conducted to determine and to measure the geometric properties of the wave. By accounting the curvature of spherical Sun, the distances of wave front from the origin as a function of time are determined. The flare is assumed to coincides with the origin of the wave.

Type II and III solar radio burst of the X-class flare of July 6 are recorded in Culgoora Observatory. While the other event is beyond the observation window of the observatory. Type III burst started to occur at 23.05 UT, while the type II radio burst started at 23.10 UT. Frequency analysis of the type II burst yields the uprising velocity of about 1200 km/s.

\section{Result and Discussion}

\begin{figure}
\centering
\includegraphics[width=0.9\textwidth]{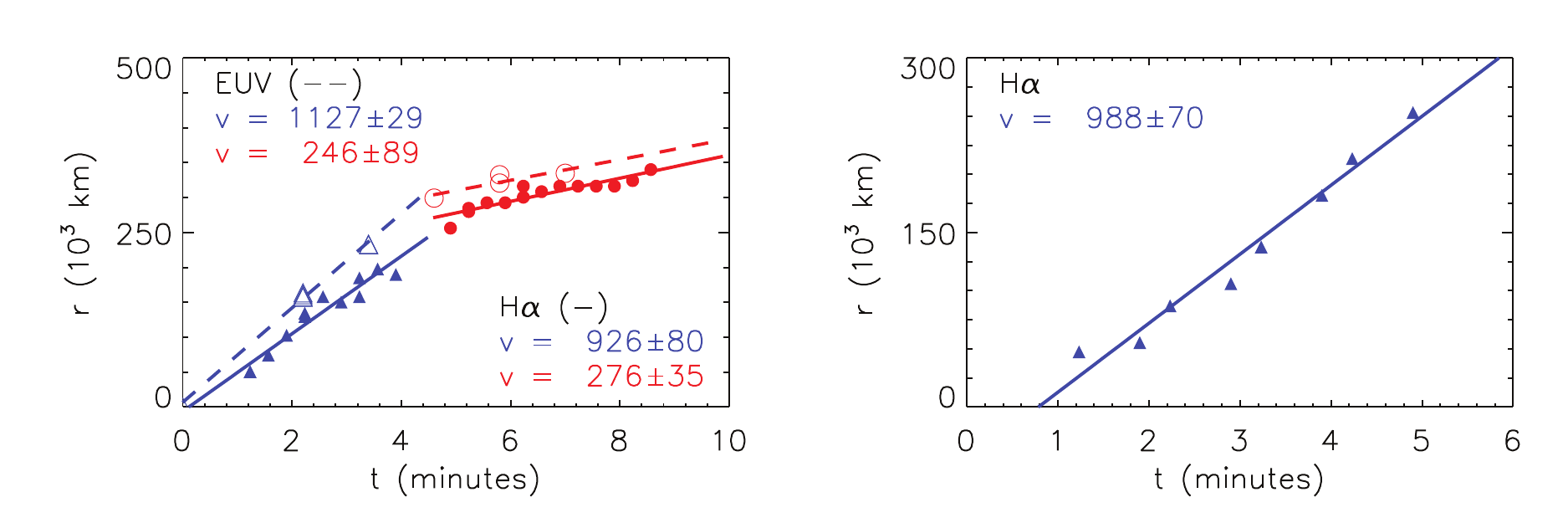}
\caption{\emph{Left}: linear speed of the shock waves of June 3, obtained from H$\alpha$ images (filled symbols) and EUV images (empty symbols). \emph{Right}: propagation speed of Moreton wave occured in July 6.}
\label{speed}
\end{figure}

Using the obtained images the velocity of the Moreton wave for the June 3 event was found to be about 926 km/sec. The same analysis was conducted for the EIT images from SDO/AIA in which the cadence is about 72 seconds, acquired the speed of the shock wave of 1127 km/sec. This resulting velocity of the EIT wave exceeds the velocity of the Moreton wave. It is not trivial since the common EIT waves have much lower velocity \citep{klassen2000}. The nature of this observed moving features (both in chromosphere and lower corona layers) are still in question. We can not confidently ensure the true oscillation of the Moreton wave without the availability of blue-wing and red-wing images of the chromosphere.

For the event of July 6, the obtained velocity is 988 km/sec and the flare was accompanied by type III burst that occurred at 23.05, about 4 minutes late compared to the peak-time of the x-ray flare. However, the radio burst time was still inside the range of abrupt increase of x-ray intensity. Type II radio burst occurred at 23.10 after the the ejected material reach higher and denser region of corona.

\bibliographystyle{aipproc}
\bibliography{paper}

\end{document}